\documentclass[fleqn,11pt,twoside]{article}
\usepackage{amsthm}
\usepackage{amsmath}
\usepackage{graphicx}

\def\half{\mbox{$\frac{1}{2}$}}
\def\d{\mbox{\rm d}}
\def\e{\mbox{\rm e}}

\makeatletter

\newcommand{\Name}[1]{\begin{flushleft}
                       \LARGE \bf #1
                       \end{flushleft}\vspace{-3mm}}

\newcommand{\Author}[1]{\begin{flushleft}
                       \it #1 \end{flushleft}}

\newcommand{\Address}[1]{\begin{flushleft}
                       \it #1 \end{flushleft}}

\newcommand{\FirstPageHead}[5]{
\begin{flushleft}
\raisebox{8mm}[0pt][0pt]
{\footnotesize \sf
\parbox{150mm}{ \qquad
 #1 #2 #3
#4\hfill {\sc #5}}}\vspace{-13mm}
\end{flushleft}}

\newcommand{\evenhead}{Author \ name}
\newcommand{\oddhead}{Article \ name}

\renewcommand{\@evenhead}{
\hspace*{-3pt}\raisebox{-15pt}[\headheight][0pt]{\vbox{\hbox to \textwidth
{\thepage \hfil \evenhead}\vskip4pt \hrule}}}
\renewcommand{\@oddhead}{
\hspace*{-3pt}\raisebox{-15pt}[\headheight][0pt]{\vbox{\hbox to \textwidth
{\oddhead \hfil \thepage}\vskip4pt\hrule}}}
\renewcommand{\@evenfoot}{}
\renewcommand{\@oddfoot}{}

\long\def\@makecaption#1#2{%
  \vskip\abovecaptionskip
  \sbox\@tempboxa{\small \textbf{#1.}\ \ #2}%
  \ifdim \wd\@tempboxa >\hsize
    {\small \textbf{#1.}\ \ #2}\par
  \else
    \global \@minipagefalse
    \hb@xt@\hsize{\hfil\box\@tempboxa\hfil}%
  \fi
  \vskip\belowcaptionskip}

\newcommand{\JNMPnumberwithin}[3][\arabic]{%
 \@ifundefined{c@#2}{\@nocounterr{#2}}{%
    \@ifundefined{c@#3}{\@nocnterr{#3}}{%
      \@addtoreset{#2}{#3}%
       \@xp\xdef\csname the#2\endcsname{%
       \@xp\@nx\csname the#3\endcsname .\@nx#1{#2}}}}%
}

\newcommand{\resetfootnoterule} {
  \renewcommand\footnoterule{%
  \kern-3\p@
  \hrule\@width.4\columnwidth
  \kern2.6\p@}
}

\renewcommand{\footnoterule}{}

\newcommand{\be}{\begin{equation}}
\newcommand{\ee}{\end{equation}}
\newcommand{\ba}{\hspace*{-5pt}\begin{array}}
\newcommand{\ea}{\end{array}}
\newcommand{\p}{\partial}

\makeatother

\numberwithin{equation}{section}
\theoremstyle{definition}

\renewcommand{\ba}{\begin{array}}
\renewcommand{\ea}{\end{array}}
\newcommand{\beg}{\begin{eqnarray}}
\newcommand{\eeq}{\end{eqnarray}}
\newcommand{\bg}{\begin{eqnarray*}}

\newcommand{\ed}{\end{eqnarray*}}
\newcommand{\n}{\newline\hfill}

\renewcommand{\p}{\partial}

\newcommand{\notlhd}{\lhd\kern-.8em{/}\ }
\newcommand{\notexist}{\ \exists\kern-.5em{\raise.1em\hbox{/}}\ }

\newcommand{\pde}[2]{\frac{\p #1}{\p #2}}

\newcommand{\inp}{{\mbox{\vbox{\hrule width0ex\hbox{\vrule
 height0ex\kern3.8pt
\vbox{\kern2.5pt}\kern3.8pt \vrule height1.6ex}
\hrule width1.6ex}}}}

\begin{document}

\renewcommand{\evenhead}{N Euler and PGL Leach}
\renewcommand{\oddhead}{Aspects of proper differential sequences}


\thispagestyle{empty}

\begin{flushleft}
\footnotesize \sf
\end{flushleft}

\FirstPageHead{\ }{\ }{\ }
{ }{{
{ }}}

\Name{Aspects of proper differential sequences of ordinary
differential equations}

\label{euler-firstpage}

\strut\hfill

\Author{N EULER and PGL LEACH\footnote{permanent address:
School of Mathematical Sciences, University of KwaZulu-Natal,
Private Bag X54001, Durban 4000, Republic of South Africa\\Email:
leachp@ukzn.ac.za; leach@math.aegean.gr}}



\Address{
Department of Mathematics,
Lule\aa\ University of Technology \\
SE-971 87 Lule\aa, Sweden\\
Email: Norbert.Euler@sm.luth.se}

\vspace{1cm}

\noindent
{\bf Abstract}:\n
We define a proper differential sequence of ordinary differential
equations and introduce a method to derive an alternative sequence 
of integrals for such a sequence. We describe some 
general properties which are illustrated by several examples.

\vspace{1cm}


\section{Introduction}
In a recent paper M Euler and the present authors 
reported a symmetry analysis and Painlev\'e analysis
of two sequences of ordinary differential equations, namely a
Riccati sequence and a  Ermakov-Pinney
sequence \cite{Euler 07 a}.  Andriopoulos and Leach \cite {And_Leach} used the singularity analysis and explicit solution of the Riccati Sequence as a vehicle to demonstrate some specific results which can arise during the course of the singularity analysis.  Subsequently Andriopoulos {\it et al} \cite {Andriopoulos 07 a} made a detailed study of the symmetry and singularity properties of the Riccati Sequence.

The aim of the present paper is to define the {\it proper differential sequence} and discuss its integrability. We also introduce 
an {\it alternative sequence} where the equations of the higher order members in the sequence do not increase in 
order, but are fixed by the first equation in the sequence. This
approach can help to integrate the sequence and provides in some cases
a direct route to the first integrals of the equations in the differential sequence.
 
The paper is organised as follows: In Section 2 we give 
definitions regarding proper differential sequences,
their Lie symmetrey algebra and their integrability. In Section 3 
we introduce a method to derive an alternative 
sequence and provide several examples to illustrate the concept
of {\it compatible} and {\it completely compatible sequences}. 
In an Appendix we provide details of the Lie point symmetry analysis of some 
of the sequences discussed in Section 3.

\section{General description}

\noindent
Consider the variables $x$ and $u$, where $u=u(x)$ with
$u_x=du/dx,\ u_{xx}=d^2u/dx^2$ and the additional notation
$u_{nx}=d^nu/dx^n$.
We now consider a differential sequence of $m$ equations,
\begin{gather}
\label{E_seq}
\{E_1,\, E_2,\,\ldots,\,E_m\},
\end{gather}
in the following form:
\begin{gather}
E_1:= F(u,u_x,u_{xx},\ldots, u_{nx})=0\nonumber\\[0.3cm]
E_2:=R^{[k]}[u]\,F(u,u_x,u_{xx},\ldots, u_{nx})=0\nonumber\\[0.3cm]
\label{seq_1}
E_3:=(R^{[k]}[u])^2\,F(u,u_x,u_{xx},\ldots, u_{nx})=0\\[0.3cm]
\qquad\qquad\vdots\nonumber\\[0.3cm]
E_m:=(R^{[k]}[u])^{m-1}\,F(u,u_x,u_{xx},\ldots, u_{nx})=0,\nonumber
\end{gather}
where
$R^{[k]}[u]$ is a $k$th-order integrodifferential
operator of the form
\begin{gather}
\label{R}
R^{[k]}[u]=G_kD_x^k+G_{k-1}D_x^{k-1}+\cdots
+G_0+Q D_x^{-1}\circ \,J.
\end{gather}
The adjoint of $R^{[k]}[u]$ has the form
\begin{gather}
\label{R^*}
(R^{[k]})^*[u]=\sum_{i=0}^k (-1)^iD_x^i\circ G_i
-JD_x^{-1}\circ Q.
\end{gather}
We term $E_1$ the {\it seed equation} of the differential sequence (\ref{seq_1}).
Note that the second equation,
\begin{gather}
E_2:=R^{[k]}[u]F(u,u_x,u_{xx},\ldots, u_{nx})=0,
\end{gather}
is of order $n+k$, the third equation,
\begin{gather}
E_3:=(R^{[k]})^2[u]F(u,u_x,u_{xx},\ldots, u_{nx})=0,
\end{gather}
is of order $n+2k$ and the $m$th equation $E_m$ is of
order $n+(m-1)k$.

\strut\hfill

\noindent
Let $L_{E_i}[u]$ denote the linear operator
\begin{gather}
L_{E_i}[u]=\pde{E_i}{u}+\pde{E_i}{u_x}D_x+\pde{E_i}{u_{2x}}D_x^2+
\ldots \pde{E_i}{u_{qx}}D_x^q
\end{gather}
and $L^*_{E_i[u]}$ the adjoint to $L_{E_i}[u]$, namely
\begin{gather}
L^*_{E_i}[u]=\pde{E_i}{u}-D_x\circ \pde{E_i}{u_x}+D_x^2\circ
\pde{E_i}{u_{2x}}+
\ldots +(-1)^qD_x^q\circ \pde{E_i}{u_{qx}}.
\end{gather}

\strut\hfill

\strut\hfill

We denote by $Z^i(E_i)$ the vertical symmetry generator
of the equation $E_i$ in the sequence (\ref{seq_1}),
namely
\begin{gather}
\label{Z^i}
Z^i(E_i)=Q(x,u,u_x,u_{xx}, u_{3x},\ldots,u_{jx})\partial_u
\end{gather}
where the necessary and sufficient invariance condition for equation $E_i$
is
\begin{gather}
\left.\vphantom{\frac{da}{db}}
L_{E_i}Q\right|_{E_i=0}=0.
\end{gather}

\noindent
Note that (\ref{Z^i}) includes the point symmetry generators
\begin{gather}
\label{nonvertical}
\Gamma_i=\xi(x,t,u)\partial_{x}+\eta(x,t,u)\partial_{u}
\end{gather}
with symmetry characteristic
$Q(x,u,u_x)=\xi(x,u) u_x-\eta(x,u)$ and equivalent vertical
form 
\begin{gather}
Z^i=\left[\xi(x,u) u_x-\eta(x,u)\right]\partial_{u}.
\end{gather}

\strut\hfill

\noindent
{\bf Definition 1:} {\it
The sequence (\ref{seq_1}) admits a $p$-dimensional Lie symmetry
algebra, ${\cal L}$, spanned by the linearly independent
symmetry generators
\begin{gather}
\{Z^i_1(E_i),\,Z^i_2(E_i),\,\ldots,\, Z^i_p(E_i)\}
\end{gather}
if each equation in the sequence (\ref{seq_1}),
$\{E_1,\,E_2,\ldots,\, E_m\}$, admits a $p$-dimensional Lie symmetry algebra,
${\cal L'}$, isomorphic to ${\cal L}$.
}

\strut\hfill

\noindent
{\bf Definition 2:} {\it
$J=J(x,u,u_x,u_{xx},\ldots)$ is an
integrating factor for the differential sequence (\ref{seq_1}) if $J$
is an integrating factor for each equation in the sequence.}

\strut\hfill

\noindent
{\bf Definition 3:} {\it
The operator $R^{[k]}[u]$ of the form (\ref{R})
is defined as a $k$th-order recursion operator of the differential sequence
(\ref{seq_1}) under the following conditions:
\begin{subequations}
\begin{gather}
\label{con_1}
\left[L_{E_i}[u],\, R^{[k]}[u]\right]=0,\qquad
i=1,2,\ldots,m,\\[0.3cm]
\label{con_2}
(R^{[k]})^*[u]J_k=\alpha J_l\ \  \forall\ \  k,l=1,2,\ldots, p,
\end{gather}
\end{subequations}
where $\alpha$ is a nonzero constant, $i=1,2,\ldots m$ and $p$
is the total number of integrating factors, $J_l$, valid for all
members of the sequence. For some values of $l$, $J_l$ may be zero.}

\strut\hfill

\noindent
{\bf Definition 4:} {\it
A proper differential sequence of ordinary differential
equations is a differential sequence which admits at least
one recursion operator of the form (\ref{R}).}

\strut\hfill

\noindent
{\bf Definition 5:} {\it
An integrable differential sequence is defined as a proper
differential sequence of ordinary differential equations
for which each equation in the sequence
is integrable.}

\strut\hfill

\noindent
{\bf Remark:} By an {\it integrable} ordinary differential equation of
$n$th order, we mean an equation which admits a solution,
$u=\phi(x;c_1,\ldots, c_n)$, where $c_j$, $j = 1,\,n $ are independent arbitrary
constants. In less strict sense we require that the $n$th-order
equation admits 
$n-1$ functionally independent first integrals such that the general
solution can be expressed as a quadrature. Integrability of a
nonlinear ordinary differential equation can also be expressed in
terms of its singularity structure in the complex domain. This
is known as the {\it Painlev\'e Property} and requires that the
solutions possess only movable poles as sigularities (see for example
\cite{Conte} for some recent reviews on the Painlev\'e Property). It can be of
interest to study the Painlev\'e Property of proper
differential sequences, but this falls outside the scope of the
current paper. We refer the to the papers by Andriopoulos {\it et al}
\cite{And_Leach, Andriopoulos 07 a} in which singularity analysis is used to study the integrability
of a Riccati sequence.

\strut\hfill

\noindent
Let 
\beg
\label{SF}
E_i:=u_{qx}-f_i(x,u,u_x,u_{xx},\ldots,u_{(q-1)x})=0,
\eeq
where
\begin{gather*}
q=n+(m-1)k.
\end{gather*}
We introduce the following total derivative operator
\beg
D_{E_i}=\left.
\vphantom{\frac{x^2}{x}}
D_x\right|_{E_i=0}
=\pde{\ }{x}+\sum_{j=0}^{q-1}u_{jx}\pde{\
}{u_{(j-1)x}}+f_i\left(x,u,u_x,\ldots,u_{(q-1)x}\right)
\pde{\ }{u_{(q-1)x}}.
\eeq

\strut\hfill

\noindent
{\bf Proposition 1:} {\it
$J_s$ is an integrating factor for the sequence (\ref{seq_1})
if and only if the following conditions are satisfied:
\begin{subequations}
\begin{gather}
\label{con_3}
\left.\vphantom{\frac{dP}{dQ}}
L^*_{E_i[u]}J_s(x,u,u_x,\ldots)\right|_{E_i=0}=0,
\qquad
i=1,2,\ldots,m,\\[0.3cm]
\pde{J_s}{u_{(q-2r)x}}
+\sum_{j=1}^{2r-1}
(-1)^{j-1}
\pde{\ }{u_{(q-1)x}}
\left\{
D_{E_i}^{j-1}\left(
\pde{f_i}{u_{(j+q-2r)x}}J_s\right)\right\}\nonumber\\[0.3cm]
\label{con_4}
+\pde{\ }{u_{(q-1)x}}\left(
D_{E_i}^{2r-1}J_s\right)=0, \qquad
s=1,2,\ldots,p,\quad
r=1,2,\ldots ,\left[\frac{q}{2}\right].
\end{gather}
\end{subequations}
Here $\left[\frac{q}{2}\right]$ is the largest natural number less than
or equal to
the number $\frac{q}{2}$,
$i=1,2,\ldots m$, and $p$ is the total number
of integrating factors, $J_s$, valid for all members of the sequence,
i.e. $s=1,2,\ldots, p.$ }

\strut\hfill

\noindent
{\bf Remark:} The derivation of the necessary and sufficient conditions for
integrating factors of single ordinary differential equations of
order $n$ are derived in the book of
Bluman and Anco \cite {Bluman_Anco}. Proposition 1 is a natural
extension of this result to proper differential sequences of ordinary differential
equations. Note that condition (\ref{con_3}) ensures that each $J_s$ is
an adjoint symmetry for each equation in the sequence and
(\ref{con_4}) ensures that this adjoint symmetry is an
integrating factor for each member of the sequence.

\strut\hfill

\noindent
{\bf Example 1:} We consider the seed equation
\begin{gather}
u_{xx}+u_x^2=0.
\end{gather}
with the differential operator
\begin{gather}
R[u]=D_x+u_x.
\end{gather}
This gives a proper differential sequence
\begin{gather}
\label{Riccati}
R^j[u]\left(
u_{xx}+u_x^2\right)=0
\end{gather}
with zeroth-order integrating factors
\begin{gather}
J_1(x,u)=e^{u},\qquad J_2(x,u)=xe^u.
\end{gather}
Here
\begin{gather}
R^{*}[u]e^u=0,\qquad R^{*}[u](xe^u)=-e^u.
\end{gather}

In the next section we show that the sequence (\ref{Riccati})
is an integrable sequence of ordinary differential equations and
discuss its properties.

\section{An alternative description}

It is of interest to construct an {\it alternative sequence},
\begin{gather}
\label{E-tilde_seq}
\{\tilde E_1,\,\tilde E_2,\,\ldots,\,\tilde E_m\},
\end{gather}
to (\ref{seq_1}), {\it ie} $\{E_1,\,E_2,\ldots,\,E_m\}$, namely one in which the order of the differential 
equations in the sequence (\ref{E-tilde_seq}) does not increase but is fixed by the seed
equation $\tilde E_1$. For the same seed equation, $E_1=\tilde E_1$,
the two sequences
(\ref{E_seq}) and (\ref{E-tilde_seq}) should then be {\it compatible} or
{\it completely compatible}.

\strut\hfill

\noindent
{\bf Definition 6:} {\it 
Two equations, $E_j$ and $\tilde E_j$ from the sequences (\ref{seq_1})
and (\ref{E-tilde_seq}) respectively, are called compatible if
the equations admit at least one common solution. The two equations
are called completely compatible if the general solution of 
$\tilde E_j$ gives the general solution for $E_j$.
Two sequences of $m$ equations,
$\{E_1,\,E_2,\,\ldots,\,E_m\}$ and 
$\{\tilde E_1,\,\tilde E_2,\,\ldots,\,\tilde E_m\}$ with the same
seed equation $E_1=\tilde E_1$, is called compatible if 
each equation in the sequence admits at least one common solution
between the corresponding members in the two sequences.
The sequences are called completely compatible if the general
solution of $\tilde E_j$ provides the general solution
for $E_j$ for all members of the sequence, ie for all
$j=1,2,\ldots,m$. The sequence (\ref{E-tilde_seq}) is termed
an alternative sequence to (\ref{seq_1}) if the two sequences are at
least compatible.
}

\strut\hfill

Since the order of the equations in an alternative sequence 
(\ref{E-tilde_seq}) does not increase, the equations that make up 
an alternative sequence should
define integrals for the equations in the proper differential sequence 
(\ref{E_seq}) to guarantee compatibility of its solutions. We
introduce the following

\strut\hfill

\noindent
{\bf Proposition 2:} {\it
Consider a proper differential sequence $\{E_1,\,E_2,\ldots,\,E_m\}$
with recursion operator $R^{[k]}[u]$.
An alternative sequence, $\{\tilde E_1,\,\tilde E_2,\ldots,\,\tilde E_m\}$,
of the form
\begin{subequations}
\begin{gather}
\tilde E_1:=F(u,u_xu_{xx},\ldots,u_{nx})=0\\[0.3cm]
\tilde
E_{j+1}:=F(u,u_xu_{xx},\ldots,u_{nx})=Q_j(x,u,u_x,\ldots,\omega^1,\omega^2,
\ldots,\omega^\ell; c_1,c_2,\ldots,c_s)\\[0.3cm] 
j=1,2,\ldots,m-1,\nonumber
\end{gather}
\end{subequations}
is compatible with the proper differential sequence 
$\{E_1,\,E_2,\ldots,\,E_m\}$ with
$E_1=\tilde E_1$ if
\begin{subequations}
\begin{gather}
R^{[k]}Q_1=0\\[0.3cm]
R^{[k]}Q_i=Q_{i-1},\qquad i=2,3,\ldots,m. 
\end{gather}
\end{subequations}
Here $\omega^1,\ \omega^2\ldots, \omega^\ell$ are nonlocal coordinates
defined by
\begin{subequations}
\begin{gather}
\frac{d\omega^1}{dx}=g_1(u),\\[0.3cm]
\frac{d\omega^2}{dx}=g_2(\omega^1),\quad
\frac{d\omega^3}{dx}=g_3(\omega^2),\ \ldots ,\
\frac{d\omega^{\ell}}{dx}=g_\ell(\omega^{\ell-1})
\end{gather}
\end{subequations}
for some differentiable functions $g_k$.
}

\strut\hfill

Below we discuss several examples of proper differential sequences
which are compatible or completely compatible
with their alternative sequences. The examples are sufficiently simple
to demonstrate the method of construction and to investigate the 
properties of the sequences.
A classification is not at all attempted here, but will be addressed in 
future works. For our examples we consider the following four seed 
equations which are part of the list of second-order
linearisable evolution equations in $(1+1)$ dimensions reported
in \cite{Euler 03 a}: 
\begin{gather*}
u_{xx}+u_x^2=0\\[0.3cm]
u_{xx}+h(u)u_x^2=0\\[0.3cm]
u_{xx}+\lambda u_x-\frac{h'(u)}{h(u)}u_x^2+h(u) = 0\\[0.3cm]
u_{xx}+uu_x=0
\end{gather*}
for arbitrary differentiable functions $h$, where prime denotes the
derivative with respect to $u$.

\noindent
{\bf Example 2:} Firstly we consider the proper differential sequence,
(\ref{Riccati}), already introduced in Example 1, where
\begin{gather*}
R[u]=D_x+u_x.
\end{gather*}
The proper differential sequence is
\begin{subequations}
\begin{gather}
\label{r_1}
E_1:=F(u,u_xu_{xx})=u_{xx}+u_x^2=0\\[0.3cm]
\label{r_2}
E_2:=R[u]F(u,u_xu_{xx})=u_{3x}+3u_xu_{xx}+u_x^3=0\\[0.3cm]
\label{r_3}
E_3:=R^2[u]F(u,u_xu_{xx})=u_{4x}+4u_xu_{3x}+3u_{xx}^2+6u_x^2u_{xx}+u_x^4
=0\\[0.3cm]
\qquad\qquad\vdots\nonumber\\[0.3cm]
\label{r_m}
E_m:=R^{m-1}[u]F(u,u_xu_{xx})=u_{(m+1)x}+\cdots =0.
\end{gather}
\end{subequations}
We apply Proposition 2 to calculate the alternative sequence
with seed equation (\ref{r_1}).
The second member in the alternative sequence is
\begin{gather}
u_{xx}+u_x^2=Q_1(x,u, u_x,\ldots)
\end{gather}
under the condition
\begin{gather}
\label{c_1}
R[u]Q_1(x,u, u_x,\ldots)=0.
\end{gather}
Condition (\ref{c_1}) is of the form
\begin{gather}
D_x(Q_1)=-u_xQ_1
\end{gather}
with general solution
\begin{gather}
Q_1(u,c_1)=c_1e^{-u},
\end{gather}
where $c_1$ is an arbitrary constant of integration.
Thus the second member in the alternative sequence is
\begin{gather}
u_{xx}+u_x^2=c_1e^{-u}.
\end{gather}
The third member, (\ref{r_3}), becomes
\begin{gather}
u_{xx}+u_x^2=Q_2(x,u,u_x,\ldots)
\end{gather}
in the alternative sequence under the condition
\begin{gather}
R[u]Q_2(x,u;c_1,c_2)=Q_1(u;c_1),
\end{gather}
which admits the general solution
\begin{gather}
Q_2(x,u;c_1,c_2)=c_1xe^{-u}+c_2e^{-u}
\end{gather}
with $c_2$ another constant of integration.
Thus the third member in the alternative sequence is
\begin{gather}
u_{xx}+u_x^2=e^{-u}\left(c_1x+c_2\right)
\end{gather}
which can be presented in the form
\begin{gather}
u_{xx}+u_x^2=e^{-u}D_x^{-1}c_1.
\end{gather}
The next member in the alternative sequence is
\begin{gather}
u_{xx}+u_x^2=e^{-u}\left(\frac{1}{2}c_1x^2+c_2x+c_3\right)\equiv
e^{-u}D_x^{-2}c_1.
\end{gather}
The functions $Q_k$ in $E_{k+1}$ are
\begin{gather}
Q_k=e^{-u}q(x)\ \ {\mbox{with}}\ \ q^{(k)}(x)=0\quad
\Leftrightarrow\quad q(x)=\sum_{j=1}^k \frac{c_j}{(k-j)!}\, x^{k-j}.
\end{gather}
\noindent
We thus conclude that
the alternative sequence to the differential sequence (\ref{r_1}) - (\ref{r_m})
has the form
\begin{subequations}
\begin{gather}
\tilde E_1:=u_{xx}+u_x^2=0\\
\tilde E_j:=u_{xx}+u_x^2=e^{-u}D_x^{-(j-2)}c_1
,\qquad j=2,3,\ldots, m,
\end{gather}
\end{subequations}
where $D_x^{-q}$ are $q\in {\cal N}$ compositions of
the integral operator $D_x^{-1}$ and $D_x^0:=1$. In explicit
form the alternative sequence is
\begin{subequations}
\begin{gather}
\label{tilde-r_1}
\tilde E_1:=u_{xx}+u_x^2=0\\[0.3cm]
\tilde E_2:=u_{xx}+u_x^2=Q_1\ \ {\mbox{with}}\ \   Q_1=e^{-u}c_1\\[0.3cm]
\tilde E_3:=u_{xx}+u_x^2=Q_2 \ \ {\mbox{with}}\ \
Q_2=e^{-u}\left(c_1x+c_2\right)
\\[0.3cm]
\tilde E_4:=u_{xx}+u_x^2=Q_3, \ \ {\mbox{with}}\ \ Q_3=e^{-u}
\left(\frac{1}{2}c_1x^2+c_2x+c_3\right) \\[0.3cm]
\ \ \vdots\\
\label{tilde-r_m}
\tilde E_m:=u_{xx}+u_x^2=Q_{m-1}
 \ \ {\mbox{with}}\ \ Q_{m-1}
=e^{-u}\left(\sum_{j=1}^{m-1}\frac{c_j}{(m-j-1)!}x^{m-j-1}\right).
\end{gather}
\end{subequations}

\noindent
It is easy to establish that the proper differential sequence
(\ref{r_1}) -- (\ref{r_m}) is an integrable differential sequence
since each member of the sequence is linearisable by the change for variables
\begin{gather}
w(X)=u_xe^{u},\qquad X=x.
\end{gather}
Moreover the alternative sequence
(\ref{tilde-r_1}) -- (\ref{tilde-r_m}) is linearisable by the change
of variables
\begin{gather}
w(X)=e^{u},\qquad X=x.
\end{gather}
To establish the compatibility or complete compatibility
of the two sequences (\ref{r_1}) -- (\ref{r_m}) and
(\ref{tilde-r_1}) -- (\ref{tilde-r_m}) we take a closer look
at the corresponding members.
\begin{itemize}
\item
Compare the members $E_2$ and $\tilde E_2$:\n
A first integral for $E_2$ is given by
$\tilde E_2$, namely
\begin{gather}
c_1=e^u\left(u_{xx}+u_x^2\right).
\end{gather}
Therefore the general solution of $\tilde E_2$ gives the general
solution of $E_2$ with $c_1$ as one of the constants of integration
for $E_2$. Hence the two equations, $E_2$ and $\tilde E_2$, are
completely compatible.
\item
Compare the members $E_3$ and $\tilde E_3$:\n
A second integral for $E_3$ is given by $\tilde E_3$, namely
\begin{gather}
c_1x+c_2=e^u\left(u_{xx}+u_x^2\right).
\end{gather}
Therefore the general solution of $\tilde E_3$ gives the general
solution of $E_3$ (with $c_1$ and $c_2$ as two of the constants of
integration for $E_3$) and the two equations $E_3$ and $\tilde E_3$ are
completely compatible. A similar argumant follows for all
equations in the proper differential sequence (\ref{r_1}) --
(\ref{r_m}).
\end{itemize}

\noindent
{\it We conclude that the two sequences  (\ref{r_1}) -- (\ref{r_m}) 
and (\ref{tilde-r_1}) -- (\ref{tilde-r_m}) are completely compatible.}

\strut\hfill

Another interesting property of the sequences
(\ref{r_1}) -- (\ref{r_m}) and (\ref{tilde-r_1}) -- (\ref{tilde-r_m})
is that the symmetry characteristics, $\eta_j$,
of the solution symmetries,
\begin{gather}
\Gamma_j^s=\eta_j(x,u)\partial_u,
\end{gather} 
for the equations in (\ref{r_1}) -- (\ref{r_m}) are given by the
functions $Q_1, Q_2,\ldots$ in the alternative sequence
(\ref{tilde-r_1}) -- (\ref{tilde-r_m}). In particular

\strut\hfill

\noindent
{\it The symmetry
characteristic, $\eta_j$, for the solution symmetry of $E_j$
in (\ref{r_1}) -- (\ref{r_m})
is given by $Q_{j+1}$ of the equation $\tilde E_{j+2}$ in
(\ref{tilde-r_1}) -- (\ref{tilde-r_m})
for all $j=1,2,\ldots, m$.}

\strut\hfill

\noindent
For example $E_1:= u_{xx}+u_x^2=0$ admits the solution symmetry
\begin{gather}
\Gamma_1^s=Q_2\partial_u,
\end{gather} 
where $Q_2=e^{-u}(c_1x+c_2)$ corresponds to the right-hand
expression in $\tilde E_3$. 

\strut\hfill

Note that the complete set of all point symmetries
for the proper differential sequence (\ref{r_1}) -- (\ref{r_m})
is the following:\n\n
For $E_1$ the complete set of Lie point symmetries are
\begin{gather}
\left\{\e ^ {-u}\partial_u,\,x\e ^ {-u}\partial_u,\,\partial_u,\,\partial_x,\,x\partial_+\half\partial_u,\,x ^ 2\partial_x+x\partial_u,\,\e ^u\partial_x,\,x\e ^u\partial_x+\e ^u\partial_u\right\}
\end{gather}
For $E_k$, $k=2,3,\ldots, m$, the complete set of Lie point symmetries are
\begin{gather}
\left\{q (x)\e ^ {-u}\partial_u,\,\partial_u,\,\partial_x,\,x\partial_x+\half (n -1)\partial_u,\,x ^ 2\partial_x+ (n -1)x\partial_u\right\},
\end{gather}
where $n$ is the order of the differential equation, $E_k$, and
\begin{gather}
q^{(n)}(x)=0.
\end{gather} 
The Lie point symmetry properties of the alternative
sequence (\ref{tilde-r_1}) -- (\ref{tilde-r_m}) are discussed
in the Appendix.

\strut\hfill

\noindent
{\bf Example 3:} Consider the seed equation
\begin{gather}
u_{xx}+h(u)u_x^2=0
\end{gather}
with recursion operator
\begin{gather}
R[u]=D_x+h(u)u_x.
\end{gather}
This defines the proper differential sequence of the form
\begin{subequations}
\begin{gather}
\label{ex2_a}
E_1:=u_{xx}+h(u)u_x^2=0\\
\label{ex2_b}
E_{j+1}:=R^j[u]\left(
u_{xx}+h(u)u_x^2\right)=0,\qquad j=1,2,\ldots, m-1.
\end{gather}
\end{subequations}
We apply Proposition 2 and calculate the functions $Q_j$ in
the same manner as in Example 2. This
leads to the following alternative sequence
\begin{subequations}
\begin{gather}
\label{tilde_ex2_a}
\tilde E_1:=u_{xx}+u_x^2=0\\
\label{tilde_ex2_b}
\tilde E_{j+1}:´= u_{xx}+u_x^2= \exp\left[-\int h(u)\d u\right]D_x^{-(j-1)}c_1
,\qquad j=1,2,\ldots, m-1.
\end{gather}
\end{subequations}
It is easy to show that the sequence (\ref{ex2_a}) --  (\ref{ex2_b})
and its alternative
sequence (\ref{tilde_ex2_a}) -- (\ref{tilde_ex2_b})  
are completely compatible and that (\ref{ex2_a}) --  (\ref{ex2_b}) 
is an integrable sequence. The linearisation and Lie point symmetries
of (\ref{tilde_ex2_a}) -- (\ref{tilde_ex2_b}) are discussed in the
Appendix.

 \strut\hfill

\noindent
{\bf Example 4:} Consider the seed equation
\begin{gather}
u_{xx}+\lambda u_x-\frac{h'(u)}{h(u)}u_x^2+h(u)=0
\end{gather}
with the recursion operator
\begin{gather}
R[u]=D_x-\frac{h'(u)}{h(u)}u_x.
\end{gather}
This gives the proper differential sequence
\begin{subequations}
\begin{gather}
\label{ex4_a}
E_1:=u_{xx}+\lambda u_x-\frac{h'(u)}{h(u)}u_x^2+h(u)=0\\[0.3cm]
\label{ex4_b}
E_{j+1}:=R^j[u]\left(u_{xx}+\lambda
u_x-\frac{h'(u)}{h(u)}u_x^2+h(u)\right)=0,
\quad j=1,2,\ldots, m-1,
\end{gather}
\end{subequations}
with its alternative sequence
\begin{subequations}
\begin{gather}
\label{tilde_ex4_a}
\tilde E_1:=u_{xx}+\lambda
u_x-\frac{h'(u)}{h(u)}u_x^2+h(u)=0\\[0.3cm]
\label{tilde_ex4_b}
\tilde E_{j+1}:=u_{xx}+\lambda
u_x-\frac{h'(u)}{h(u)}u_x^2+h(u)=h(u)D_x^{(j-1)}c_1,
\quad j=1,2,\ldots, m-1,
\end{gather}
\end{subequations}
where $h$ is an arbitrary differentiable function. Just like the
sequences in Example 2 and Example 3 the sequences 
(\ref{ex4_a}) -- (\ref{ex4_b}) and (\ref{tilde_ex4_a}) -- (\ref{tilde_ex4_b})
are completely compatible and (\ref{ex4_a}) -- (\ref{ex4_b}) is an
integrable sequence. Details are given in the Appendix.

\strut\hfill

\strut\hfill

\noindent
{\bf Example 5:}
To the Burgers equation
\begin {equation}
u_{xx} + uu_x = u_t \label {3.161}
\end {equation}
one can (with standard symmetry reduction following the
$t$-translation invariance)
associate
\begin {equation}
u_{xx} + uu_x = 0 \label {3.16}
\end {equation}
which shares the same integrodifferential recursion operator,
\begin {equation}
R [u ] = D_x+ \half u+ \half u_xD_x ^ {-1}\circ 1. \label {3.17}
\end {equation}
The proper differential sequence, which we name the {\it Burgers Sequence},
is
\begin{subequations}
\begin{gather}
\label{ex5_a}
E_1:=u_{xx} + uu_x = 0\\
\label{ex5_b}
E_{j+1}:=R^j[u]\left(u_{xx} + uu_x\right)=0,\qquad j=1,2,\ldots,m.
\end{gather}
\end{subequations}
We now construct an alternative Burger's Sequance following Proposition 2.
The solution of $R [u ]Q_1 = 0 $ is
\begin {equation}
Q_1 = \left (- 2 A\exp\left [ -\half\int u\d x \right] + 2 B\exp\left [ -\half\int u\d x \right]\int\exp\left [ \half\int u\d x\right] \d x\right)_{x}, \label {3.18}
\end {equation}
where $A $ and $B $ are constants of integration.  Equation (\ref
{3.18}) is 
an integrodifferential equation.  It can be rendered as an ordinary differential equation by defining
\begin {equation}
w = \int\exp\left [ \half\int u\d x \right]\d x \label {3.19}
\end {equation}
so that
\begin {equation}
u_{xx} + uu_x = \left (- 2 A\exp\left [ -\half\int u\d x \right] + 
2 B\exp\left [ -\half\int u\d x \right]\int\exp\left [ 
\half\int u\d x \right]\d x\right)_{x} \label {3.20}
\end {equation}
becomes
\begin {equation}
\frac {w_{4x}} {w_{x}} -\frac {w_{xx}w_{3x}} {w_{x} {} ^ 2} = \frac {Aw_{xx}} {w_{x} {} ^ 2} + B\left (1 - \frac {ww_{xx}} {w_{x} {} ^ 2}\right). \label {3.21}
\end {equation}

\strut\hfill

\noindent
In a similar fashion the equation $R [u ]Q_2 = Q_1 $ has the solution
\begin {gather}
Q_2 = \left\{2 C \exp
\left [ -\half\int u\d x \right]
\int\exp\left[ \half\int u\d x \right]\d x
- 2 Ax\exp\left [ -\half\int u\d x \right]
\right.
\nonumber \\[0.3cm]
\left.\qquad + 2
B\exp\left [ -\half\int u\d x \right]
\int\left (\int\exp\left [ \half\int u\d x \right]\d x\right)
\d x\right\}_{x}, \label {3.22}
\end {gather}
where $C $ is also a constant of integration, and the
integrodifferential
equation is
\begin {gather}
u_{xx} + uu_x = \left\{2 C \exp\left [ -\half\int u\d x
\right]\int\exp\left [ \half\int u\d x \right]\d x- 2Ax\exp\left [
-\half\int u\d x \right] \right.\nonumber \\[0.3cm]
\left.
\qquad + 2 B\exp\left [ -\half\int u\d x \right]\int\left(
\int\exp\left [ \half\int u\d x \right]\d x\right)\d x\right\}_{x}.
\label {3.23}
\end {gather}
The corresponding higher-order ordinary differential equation is
\begin {equation}
\frac {w_{5x}} {w_{xx}} -\frac {w_{3x}w_{4x}} {w_{xx} {} ^ 2} = C\left ( 1 -\frac {w_{x}w_{3x}} {w_{xx} {} ^ 2}\right) + A\left (\frac {xw_{3x}} {w_{xx} {} ^ 2} -\frac {1} {w_{xx}}\right) + B\left (\frac {w_{x}} {w_{xx}} - \frac {ww_{3x}} {w_{xx} {} ^ 2}\right), \label {3.24}
\end {equation}
where now
\begin {equation}
w = \int\left (\int\exp\left [ \half\int u\d x \right]\d
x\right)\d x \label {3.25}
\end {equation}
or equivalently
\begin{gather}
\label{w_transform}
u=2\frac{w_{3x}}{w_{xx}}.
\end{gather}


Evidently one could continue in like fashion.  It contrast to 
Examples 2, 3 and 4 in which the recursion operator did not contain an inverse
derivative to obtain an ordinary differential equation one must
redefine
the dependent variable as in (\ref {3.19}) and (\ref {3.24}) (and the
obvious extension for higher elements of the sequence).  We note that
the terms on the left sides of (\ref {3.21}) and (\ref {3.23}) have
the
same form apart from the increase in the order of each derivative.
Consequently, if we wish to have a differential sequence based upon
the
differential equation (\ref {3.16}) and its recursion operator,
(\ref {3.17}), of $m $ elements in terms of differential equations,
all differential equations belonging to the sequence must be written
in terms of a differential equation of order $m+2 $.  The alternative
is an integrodifferential equation of increasing nonlocality.

\strut\hfill

We thus conclude that the first three terms in the alternative sequence take the
following forms
%
\begin {subequations}
\begin {gather} 
\label {3.26_a}
\tilde E_1(w):=\frac {w_{5x}} {w_{xx}} -\frac {w_{3x}w_{4x}} {w_{xx}
  {} ^ 2} = 0 \\[0.3cm]
\qquad\quad \Leftrightarrow\ 
\left(\frac{w_{4x}}{w_{xx}}\right)_x=0\\[0.3cm]
\qquad\quad\Leftrightarrow\quad 
w_{4x}=k_1 w_{2x}\\[0.3cm]
\label {3.26_b}
\tilde E_2(w):=\frac {w_{5x}} {w_{xx}} -\frac {w_{3x}w_{4x}} {w_{xx} {} ^ 2} 
= \frac {Aw_{3x}} {w_{xx} {} ^ 2} + B\left (1 - 
\frac {w_{x}w_{3x}} {w_{xx} {} ^ 2}\right) \\[0.3cm]
\qquad\quad \Leftrightarrow\ 
\left(\frac{w_{4x}}{w_{xx}}\right)_x
=-\left(\frac{A}{w_{xx}}\right)_x+B\left(\frac{w_x}{w_{xx}}\right)_x\\[0.3cm]
\label{tilde_E_2(w)}
\qquad\quad \Leftrightarrow\ 
w_{4x}=a_1w_{xx}+Bw_x-A\\[0.3cm]
\tilde E_3(w):=\frac {w_{5x}} {w_{xx}} -\frac {w_{3x}w_{4x}} {w_{xx} {} ^ 2} 
= C\left ( 1 -\frac {w_{x}w_{3x}} {w_{xx} {} ^ 2}\right) 
+ A\left (\frac {xw_{3x}} {w_{xx} {} ^ 2} -\frac {1} {w_{xx}}\right)\nonumber\\[0.3cm]
\label {3.26_c}
\qquad\quad\qquad\quad \qquad\quad \qquad  
+ B\left (\frac {w_{x}} {w_{xx}} - \frac {ww_{3x}} {w_{xx} {} ^
    2}\right)\\[0.3cm]
\qquad\quad \Leftrightarrow\ 
\left(\frac{w_{4x}}{w_{xx}}\right)_x
=C\left(\frac{w_x}{w_{xx}}\right)_x-A\left(\frac{x}{w_{xx}}\right)_x
+B\left(\frac{w}{w_{xx}}\right)_x\\[0.3cm]
\label{tilde_E_3(w)}
\qquad\quad \Leftrightarrow\ 
w_{4x}=a_2w_{xx}+Cw_x+Bw-Ax.
\end {gather}
\end {subequations}
However, such a differential sequence should not be confused 
with the normal type of differential sequence since, as the value of 
$m $ increases, both the left side of the equations and the recursion 
operator must be redefined.

In order to establish compatibility of the first three members of the
two sequences, (\ref{ex5_a}) -- (\ref{ex5_b}) and 
(\ref{3.26_a}) -- (\ref{3.26_c}), need to be written in the same
variable $w$, that is, we need to apply the transformation
(\ref{w_transform}) and transform the first three members of the
differentrial sequence (\ref{ex5_a}) -- (\ref{ex5_b}) in order to
write the equations in terms of the variable $w$.
We obtain the following:
\begin{subequations}
\begin{gather} 
E_1(w):=\frac {w_{5x}} {w_{xx}} -\frac {w_{3x}w_{4x}} {w_{xx} {} ^ 2}
= 0
\ \Leftrightarrow\ 
\left(\frac{w_{4x}}{w_{xx}}\right)_x=0\quad \Leftrightarrow\quad 
w_{3x}=k_1 w_{x}+k_{11}
\label{E_1(w)}
\\[0.3cm]
\label{E_2(w)}
E_2(w):=\frac {w_{6x}} {w_{xx}} -\frac {w_{3x}w_{5x}} {w_{xx} {} ^ 2}
= 0
\ \Leftrightarrow\ 
\left(\frac{w_{5x}}{w_{xx}}\right)_x=0\quad \Leftrightarrow\quad 
w_{4x}=k_2 w_{x}+k_{21}
\\[0.3cm]
\label{E_3(w)}
E_3(w):=\frac {w_{7x}} {w_{xx}} -\frac {w_{3x}w_{6x}} {w_{xx} {} ^ 2}
= 0
\ \Leftrightarrow\ 
\left(\frac{w_{6x}}{w_{xx}}\right)_x=0\quad \Leftrightarrow\quad 
w_{5x}=k_3 w_{x}+k_{31}
\end{gather}
\end{subequations}
All $k$s are constants of integration.

\strut\hfill

\noindent
Comparing (\ref{tilde_E_2(w)}) with (\ref{E_2(w)}) and 
(\ref{tilde_E_3(w)}) with (\ref{E_3(w)}) it is clear that these two
sequences are compatible but not completely compatible. 
In particular the expression
\begin{gather}
\label{comp}
w_{4x}=Bw_x-A,
\end{gather}
which is (\ref{tilde_E_2(w)}) with $a_1=0$, is a second integral
for $E_2$, namely equation (\ref{E_2(w)}). However,
the same second integral for (\ref{E_3(w)}) follows
from (\ref{tilde_E_3(w)}), where the two additional parameters
$a_2$ and $C$ have to be zero for compatibility. Therefore
the higher members of the alternative sequence do not provide
additional parameters for the integration of the 
differential sequence (\ref{ex5_a}) -- (\ref{ex5_b}) and we conclude
that the two sequences only share special solutions.
The proper differential sequence,
(\ref{ex5_a}) -- (\ref{ex5_b}), is, however, integrable since every
member of the sequence can be linearised. The same is of course true
for the alternate sequence (\ref{3.26_a}), (\ref{3.26_b}) and (\ref{3.26_c}).

\strut\hfill

Unlike the Examples 2,3 and 4 there is no preservation of Lie
point symmetries in the alternate sequence.  In the cases of 
(\ref {3.26_a}), (\ref {3.26_b})
and  (\ref {3.26_c}) we obtain
\begin {subequations}
\begin {gather} \label {3.27}
\Gamma_1 = \partial_{x},\,\Gamma_2 = \partial_{w},\,
\Gamma_3 = x\partial_{x},\,
\Gamma_4 = x\partial_{w},\,
\Gamma_5 = w\partial_{w} \\[0.3cm]
\Gamma_1 = \partial_{x},\,
\Gamma_2 = \partial_{w},\,
\Gamma_6 = (Ax -Bw)\partial_{w} \\[0.3cm]
\Gamma_7 = B\partial_{x}+ A\partial_{w},\,
\Gamma_8 = (ABx - B ^ 2w - AC)\partial_{w},
\end {gather}
\end {subequations}
respectively.

\strut\hfill

Above we only looked at the first three members of the sequence. We
end this example with the following statements about the full sequence
(\ref{ex5_a}) -- (\ref{ex5_b}).

\strut\hfill

\noindent
{\it
The $n $th element of the alternative Burgers Differential Sequence 
written in the integro-\n
differential form
\begin {equation}
u_{xx} + uu_x = \exp\left[ -\half\int u\d x\right]
\left(\sum_{i = 1} ^ {n-1}B_iD_x ^ {-i}\exp
\left[\half\int u\d x\right]\right) \label {3.43}
\end {equation}
is linearised to
\begin {equation}
W_{(n+1)x} = B_{n-2} + B_{n -1}W, \label {3.44}
\end {equation}
where $W = D_x ^ {-(n -1)}\exp\left [\half\int u\d x\right ]$.}

\strut\hfill

\noindent {\bf Remark:}  This procedure for the linearisation of (\ref
{3.43}) is a 
natural generalisation of the Cole-Hopf transformation which can also
be derived via the $x$-generalised hodograph transformation for
evolution equations \cite{Euler 03 a}.

\strut\hfill

\noindent
{\it The $n $th element of the Burgers Differential Sequence
  (\ref{ex5_a}) -- (\ref{ex5_b}), ie
\begin {equation}
R^{n -1}[u]\left (u_{xx} +uu_x\right) = 0, \label {3.45R}
\end {equation}
where $R [u ] = D_x+ \half u+ \half u_xD_x ^ {-1} $, is linearised to
\begin {equation}
v_{(n+1)} = \Omega ^ {n+1}_n v, \label {3.45}
\end {equation}
where $u = 2v_x/v $ and $\Omega$ are arbitrary constants.}

\strut\hfill

\noindent {\bf Remark:}  Here we make use of the relationship between
the elements of the Burgers Differential Sequence (\ref{ex5_a}) -- (\ref{ex5_b})
and the sequence in Example 2.

\section{Discussion}
Although no general Theorem has been provided to investigate the
integrability of a proper differential sequence, the paper
gives definitions of these objects and suggests some
roots for the investigations illustrated by several examples.
In particular, the foregoing examples strongly suggest that
the alternative sequence and Proposition 2, which addresses
the compatibility/complete
compatibility of sequences, provides a useful root to the
integrals of a proper differential sequence. In this sense, the current
paper should been appreciated  as a starting point for the investigations
of proper differential sequences rather than a concluding paper on
this subject.

We deliberately concentrate on simple examples, namely proper differential
sequences for which
the general solution can be derived via linearisations of
the equations in the sequences, in order to gain an understanding of
the properties. We recall that equations of the
sequence in the last example, the Burger' Sequence of Example 5,
are linearisable by a Cole-Hopf type transformation, whereas all
other sequences are examples of equations linearisable by point
transformations. The point-linearisable sequences have beautiful
propereties in view of the Lie symmetry structure of the sequence
and the usefulness of the alternative sequence for the construction of
the complete set of first integrals of the proper differential sequences.
For the Burgers' Sequence we have to introduce
nonlocal variables for the general solution of the operator
equation
\begin{gather}
R[u]Q_1=0
\end{gather}
which then results in a higher order alternative sequence
in terms of local variables. This example clearly suggests
that further investigations are necessary in order to handle
such cases, namely when nonlocal variables come into play. 
We suspect that nonlocal symmetries and nonlocal integrating
factors will play an important role for this investigation. 

\appendix
\section{Appendix}

The sequences discussed in Examples 2, 3 and 4
are integrable sequences and their 
alternative sequences preserve the maximal 
Lie symmetry algebra of point symmetries as given by the 
corresponding seed equations. A detailed Lie point symmetry analysis
is the aim of this Appendix.

\strut\hfill

We determine the Lie
point symmetries 
of the general equations in the systems
\begin{subequations}
\begin{gather}
\label {3.1a}
u_{xx}+u_x^2 = 0 \\
\label{3.1b}
u_{xx}+u_x^2 = e^{-u}D_x^{-(j-1)}c_1,\qquad j = 1,\, m,
\end{gather}
\end{subequations}
\begin{subequations}
\begin{gather}
\label {3.2a}
u_{xx}+u_x^2 = 0 \\
\label {3.2b}
u_{xx}+u_x^2 =
\exp\left[-\int h(u)du\right]D_x^{-(j-1)}c_1,
\qquad j = 1,\, m,
\end{gather}
\end{subequations}
and
\begin{subequations}
\begin{gather}
\label{3.3a}
u_{xx}+\lambda u_x-\frac{h'(u)}{h(u)}u_x^2+h(u) = 0\\[0.3cm]
\label{3.3b}
u_{xx}+\lambda u_x-\frac{h'(u)}{h(u)}u_x^2+h(u) =
h(u)D_x^{(j-1)}c_1.
\end{gather}
\end{subequations}
It is evident that (\ref {3.1a}) -- (\ref {3.1b})
is subsumed into (\ref {3.2a}) -- (\ref {3.2b}).

\strut\hfill

Equation (\ref {3.2b}) belongs to the class of equations
\begin {gather}
u_{xx}+u_x^2= \exp\left[-\int h(u)\d u\right] f (x) \label {3.4}
\end {gather}
and (\ref {3.3b}) to the class
\begin {gather}
u_{xx}+\lambda u_x-\frac{h'(u)}{h(u)}u_x^2+h(u) = h(u) f (x),
\label {3.5}
\end {gather}
where $f (x) $ is at least $C ^ 1 $.  This is more than adequately
covers the polynomials of the original equations.

\strut\hfill

If in (\ref {3.4}) we make the change of dependent variable,
\begin {gather}
w =\int\exp\left [\int h (u)\d u\right ]\d u, \label {3.6}
\end {gather}
(\ref {3.4}) becomes
\begin {gather}
w_{xx} = f (x). \label {3.7}
\end {gather}
In like manner the change of dependent variable,
\begin {gather}
w = \int\frac {\d u} {h (u)}, \label {3.8}
\end {gather}
converts (\ref {3.5}) to
\begin {gather}
w_{xx} + \lambda w_x = f (x) -1. \label {3.9}
\end {gather}
Since the transformation in the dependent variable is a point transformation in both cases, the Lie point symmetries of (\ref
{3.7}) and (\ref {3.9}), which are reasonably easy to calculate, lead directly to the Lie point symmetries of (\ref {3.4}) and
(\ref {3.5}).  Since (\ref {3.7}) and (\ref {3.9}) are linear second-order ordinary differential equations, they each possess
eight Lie point symmetries with the algebra $sl (3,R) $.

\strut\hfill

The coefficient functions for a Lie point symmetry of (\ref
{3.7}),
\begin{gather}
\Gamma = \xi (x,w)\partial_{x}+ \eta (x,w)\partial_{w},
\end{gather}
have the forms
\begin {subequations}
\begin {gather} \label {3.10}
\xi = a (x) + b (x)w \\
\eta = b_{x} (x)w ^ 2+ c (x)w+ d (x),
\end {gather}
\end {subequations} in which the functions of $x $ satisfy
\begin{subequations}
\begin {gather}
\label {3.11a}
b_{xx} = 0 \\
\label {3.11b}
c_{xx} = bf_{x} \\
\label {3.11c}
a_{xx} = b +2c_{x} \\
\label {3.11d}
d_{xx} = af_{x} + (2 a_{x} -c)f.
\end {gather}
\end{subequations}
We solve the equations in (\ref {3.11a}) -- (\ref{3.11d})
in turn to obtain
\begin{subequations}
\begin {gather*}
\label {3.12a}
b = B_0 + B_1x \\
\label {3.12b}
c = C_0 + C_1x + \int\int (bf_{x})dx\,dx \\
\label {3.12c}
a = A_0 + A_1x + \int\int\left (b +2c_{x}\right)dx\,dx \\
\label {3.12d}
d = D_0 + D_1x + \int\int\left [ af_{x} + (2 a_{x} - c)f\right ]dx\,dx.
\end {gather*}
\end{subequations}
We did not give the explicit formul\ae\ for the integrals for
general $f (x) $ as they are not informative.  The important thing
to note is that there are eight Lie point symmetries.

\strut\hfill

The Lie point symmetries of (\ref {3.9}) have the same dependence
upon $w $ as given in (\ref {3.10}).  Now the equations to be
satisfied by the functions of $x $ are
\begin{subequations}
\begin {gather*} \label {3.13}
b_{xx} - \lambda b_{x} = 0 \\
c_{xx} + \lambda c_{x} = bf_{x} +2\lambda b (f -1) \\
a_{xx} -\lambda a_{x}  = 2c_{x} -3b (f -1) \\
d_{xx} + \lambda d_{x} = af_{x} + (2 a_{x} -c) (f -1)
\end {gather*}
\end{subequations}
which can easily be solved and again provides eight arbitrary
constants and hence eight Lie point symmetries.

\strut\hfill

The explicit Lie point symmetries of the original equations, (\ref
{3.2b}) and (\ref {3.3b}), require both the inversion of the
transformations (\ref {3.6}) and (\ref {3.8}) and the specification of
the function $f (x) $.  To maintain a modicum of simplicity we take
(\ref {3.1b}), for which $u = \log w $, and $f (x) = c_3 + c_2x +
\half c_1x ^ 2 $, {\it ie} we list the Lie point symmetries of the
equation which is completely compatible with the fourth element of the
differential sequence (\ref{Riccati}).
\begin {gather*} \label {3.15}
\Gamma_1 = \e^{-u} \partial_{u} \\[0.3cm]
\Gamma_2 = x\e^{-u}  \partial_{u}\\
\Gamma_3 = 6 \partial_{x} + \left(3 c_2 x^2+c_1x^3\right)\e^{-u}
\partial_{u}\\[0.3cm]
\Gamma_4 = \left [ 24-\left (12 c_3 x^2 + 4 c_2 x^3 + c_1
x^4\right)e^{-u} \right ]
\partial_{u}\\[0.3cm]
\Gamma_5 = 6 x \partial_{x}+ \left(6 c_3 x^2+3 c_2 x^3+c_1
x^4\right)\e^{-u}
 \partial_{u}\\[0.3cm]
\Gamma_6 = 24 x^2\partial_{x}+ \left [24 x+\left (12 c_3 x^3+8 c_2
x^4+3 c_1 x^5\right)
\e^{-u} \right ] \partial_{u}\\[0.3cm]
\Gamma_7 = \left(144 \e^u-216 c_3x^2-24 c_2 x^3-6 c_1 x^4\right)
\partial_{x}
+ \left [ 72 c_2 x^2+24 c_1 x^3 \right. \nonumber \\[0.3cm]
\qquad \left. -\left (144 c_3^2 x^3 +108 c_3 c_2 x^4 + 12 c_2^2x^5 +
36 c_3 c_1 x^5 + 7 c_2 c_1 x^6 +c_1^2 x^7\right)\e^{-u} \right ]
\partial_{u}\\[0.3cm]
\Gamma_8 = \left(576 \e^u x-288 c_3 x^3-96 c_2 x^4-24 c_1 x^5\right)
\partial_{x}+ \left [ 576 e^u+96 c_2 x^3 \right. \nonumber \\[0.3cm]
\qquad \left. +48 c_12x^4-\left (144 c_3^2 x^4 + 144c_3 c_2 x^5 + 32
c_2^2 x^6 + 48 c_3 c_1 x^6 + 20  c_2 c_1 x^7\right.\right.\\[0.3cm]
\left.\left.\qquad+ 3 c_1^2x^8\right)\e^{-u} \right ]\partial_{u}.
\end {gather*}


\strut\hfill

\section*{Acknowledgements}
The work reported in this paper is part of a project funded under the
Swedish-South African Agreement, Grant Number 60935, sponsored by
sida/VR and the National Research Foundation of the Republic of South
Africa. PGLL thanks the Department of Mathematics, Lule\aa\ University
of Technology, for the provision of facilities while the bulk of this work was undertaken.  PGLL thanks the University of KwaZulu-Natal for its continued support.

\strut\hfill

\end{document}